\PassOptionsToPackage{hyphens}{url}

\documentclass[%
 reprint,pre,
 amsmath,amssymb,
 aps,
]{revtex4-2}

\usepackage{graphicx}
\usepackage{dcolumn}
\usepackage{bm}
\usepackage{color}
\usepackage{soul}
\usepackage[separator = {;}]{credits}
\usepackage{booktabs}
\usepackage{float}

\setstcolor{red}

\begin{document}

\preprint{APS/123-QED}

\title{Evolution of mutating pathogens in networked populations}

\author{Aviel Ivry}
\email{aviel9363@gmail.com}
 \affiliation{Department of Physics, Bar-Ilan University, Ramat-Gan 52900, Israel} 
\author{Reuven Cohen}
 \email{reuven@math.biu.ac.il}
\affiliation{Department of Mathematics, Bar-Ilan University, Ramat-Gan 52900, Israel}
\author{Amikam Patron}
 \email{patron@g.jct.ac.il}
 \affiliation{Department of Mathematics, Jerusalem College of Technology, Jerusalem 91160, Israel}

\date{\today}

\begin{abstract}

Epidemic spreading over populations networks has been an important subject of research for several decades, and especially during the Covid-19 pandemic. Most epidemic outbreaks are likely to create multiple mutations during their spreading over the population. In this paper, we study  the evolution of a pathogen which can mutate continuously during the epidemic spreading. We consider pathogens whose  mutating parameter is the mortality mean-time, and study the evolution of this parameter over the spreading process.
We use analytical methods to compute the dynamic equation of the epidemic and the conditions for it to spread. We also use numerical simulations to study the pathogen flow in this case, and to understand the mutation phenomena. We show that the natural selection leads to less violent pathogens becoming predominant in the population. We discuss a wide range of network structures and show how different effects are manifested in each case.
We also applied our theory in the context of the Covid-19 pandemic, using relevant epidemiological data collected for this outbreak. We provided explanations for the variants spreading processes observed throughout this pandemic.

\end{abstract}

\maketitle

\section{\label{sec:intro}Introduction}

Epidemic spreading has been studied thoroughly over several decades. Due to the impact that epidemics have had on the course of human history, epidemics are a subject of major interest for mathematicians, biologists, physicians and historians alike.

Classical mathematical works in the field (during the 20th century) developed some compartmental models to describe epidemics spread in population. The most common models are SIR and SIS \cite{mathTheory, math_infec}. The mean-field versions of these models are based on division of the population to groups based on their epidemic status (Susceptible/Infected/Recovered and more groups in some variations of the models) and solving ordinary differential equations to describe the population groups evolution over time, during the epidemic spreading process.

Understanding the process in a network can yield significant insights above and beyond simple mean-field approaches. Indeed, many physical, social, and biological phenomena can be well modeled by complex networks and can be studied using network principles \cite{Networks_ev, barabasiScience, PhysRevLett.85.5468, network_brain,PhysRevE.66.035101, Rocha2022_Social}. Previous literature on network modeling employs statistical methods in conjunction with algorithmic analysis to study and model the behavior and characteristics of many kinds of networks.

Accordingly, in the last two decades epidemic spreading has been intensively studied using models of networks theory \cite{Newman_spread_inNetworks, SWeffect_epidemics, TH_epidemics_net,TH_net_SIS, analysis_appl,PhysRevE.83.025102, RevModPhys.87.925, WANG2019292}.
This trend also intensified during the Covid-19 pandemic, where network based epidemic models were applied and developed, in an attempt to prevent the spread of the disease \cite{maheshwari-appnetworksc-2020,thurner-proc_acad_sc-2020}. While classical deterministic models assume a homogeneously-mixed population, in practice populations are not fully-mixed and more likely to behave as a complex network \cite{bansal-interface-2007}. Namely, in a real population diseases spread between individuals only when they are actually in contact with one-another. Therefore, when modeling epidemics one should consider the population structure. Common simplifications like considering a population-wide ``contact rate'' do not take this structure into account. Considering this structure leads to more realistic models of the epidemic.

The spreading of epidemics in networks has been studied over different network structures under different conditions and different epidemic models (which differ in the way the population is divided). Newman \cite{Newman_spread_inNetworks} studied epidemic spreading on a wide variety of networks, based on the classical SIR model. In this work, he defined a ``transmissibility'' of an epidemic, that is its ability to spread, as follows:
\begin{equation}
T:=1-\int_0^\infty dr~ d\tau~P(r)P(\tau)e^{-r\tau}\;,
\end{equation}
where $P(r)$ and $P(\tau)$ are the probability densities of the infection rate and infection duration, respectively. Newman found the critical value to reach an endemic state as:
\begin{equation}
    \label{eq:tc}
        T_c = \frac{<k>}{<k^2>-<k>} \;,
\end{equation}
where $<k>$ is the expectation of the number of links of each node in the network. The ``transmissibility'' can be also described as a general version of the transmission coefficient.

Epidemic thresholds have been studied in a wide range of cases based on different networks structures and epidemic models. For example, Parshani $et\:al.$ \cite{TH_net_SIS} used the method of percolation theory and the SIS model spreading on a random graph, to calculate the probability of reinfection. This enables them to compute the $R_0$ parameter, that is the average expected infections of each infected node. The $R_0$ used to get the critical values of an epidemic, when for an epidemic occurrence we demand $R_0>1$.

Another effect of the network's structure on epidemics spread, is the impact of the parameters and properties of the epidemic seeds (first infected nodes) of the network, on the spreading rate. The impacting parameters include not only the degree of a seed, but also the degree of its neighbors and its location in the network \cite{Kitsak2010}. Some works such as \cite{Min_B_influencers} tried to define the main spreaders in the network and calculate the correlation between the centrality of a seed and the probability of an epidemic in the SIR model. Another work \cite{PhysRevResearch.2.023332} studied the subject using analytical calculations and numerical tools in order to solve the problem of identifying influential spreaders in the SIS dynamics on generic networks. It is based on the QMF framework \cite{PhysRevE.85.066131} to get the relevant closed-form equations.

As we have seen intensely during the Covid-19 pandemic, epidemics are likely to create new mutations during their spread over the population \cite{covidMutations_2020}. As a result, we may have more than one virus/pathogen spreading in the same population at the same time, competing for dominance in the networked population. These pathogens do not have to be similar, when sometimes two different epidemics can be spread at the same time. In Ref. \cite{2pathogensNewman} the threshold for one pathogen to block the spread of another pathogen and reach dominance over the network, i.e. the ``co-existence threshold'', was calculated. In addition, the work showed that between the critical transmission value of an epidemic spreading and the critical transmission value for the first pathogen to block the spread of the second, there are intermediate transmission values where both pathogens can spread together.
While this work is focused on two pathogens that spread in succession, the other case where the two pathogens start to spread at the same time was also studied, and the condition for one pathogen to reach dominance over the other was also calculated in this case \cite{CompetingEpidemic_PRE}.

In reality, epidemics are more likely to have more than two mutations, since mutations are created randomly during the epidemic spreading. Therefore, some works have also studied cases of multi pathogen spread, or continuous mutated pathogen.

The mutation of a pathogen can be created by changing one of its parameters, such as the infection probability \cite{mutatung_by_I}. Anyway, in order to induce a competition based on a pathogen's fitness, the changed parameter must affect the ''transmissibility'' of the pathogen.

In this paper we study the continuous-time mutation of a pathogen, where the mutated parameter is the mortality mean-time of the disease it causes. We derive analytically and solve the epidemic equations for this case. We also use simulations to confirm our calculations and describe the pathogens' flow over different types of networks.


\section{\label{sec:mutating}mutating pathogens}
In the real world, infectious diseases mutate constantly. Although newly infected individuals are often contagious with the same virus as the individual they infected from, in practice each new infected individual may be infected with a somewhat different pathogen. Sometimes these little changes can create a significant difference between different pathogens that evolved from the same primary epidemic, spreading together in the same population. In such cases, one new pathogen, probably the ``fittest'' pathogen that exists in the network, may be able to gain dominance over the population and as a result to infect more susceptible individuals compared to older pathogens. A pathogen will be considered as fitter, if it has higher transmissibility. It can either be a pathogen with longer lifetime or a pathogen with higher probability to infect.

Here we assume that each secondary infection differs slightly from its infection source. For the analytical calculations as well as the numeric, we employ a standard SIR model with minor adjustment - the recovery population is also divided to recovered nodes who became healthy and immunized and recovered nodes who died due to the pandemic. Furthermore, we assume that immunity following recovery is permanent even to mutated variants, so that individuals who were infected and recovered can not get infected again even with other pathogen(unlike recent works such as \cite{PhysRevE.65.031915}). This means that all the pathogens in our studied case are from the same ``family''.

We define three main parameters to describe the pathogen -- the average time to infect a neighbor node ($\lambda$), the average time to recover and get healthy ($r$), and the average time for an infected node to die from the disease (or to get recovered in the SIR model also) ($\gamma$). The $\gamma$ parameter can also be defined as ``mortality mean time" or ``virulence". The recovery and mortality are considered to be competing Poisson processes, where the process that is activated first ``wins'' (i.e. the competition between them determines whether the node recovers or dies). In term of the spreading process, both outcomes lead to the end of the node's infectious state.

The mutation process is defined by randomly changing $\gamma$ for each newly infected node, where the change can be in both sides: the new infected node gets a new pathogen with higher or lower mortality mean time. Therefore, the mutated parameter ($\gamma$, or the mortality mean time) on new infected node shall be defined as below:
\begin{equation}
    \label{eq:mutation_process_one_node}
        \gamma_\mathrm{new} = \gamma_\mathrm{old} \xi ^a,
\end{equation}
where $a$ is randomly chosen as $a=\pm 1$ and $\xi$ is the ``mutating parameter", which determines the mortality mean time step of mutation and defined specifically for each pandemic (the mutating parameter shall be bigger than one but not too big, since big steps will change the pathogens too fast). 
Accordingly, every infected node can be described by the summation of the left and right ``$a$ steps" that occurred during the spread from the initial pathogen to it. We denote the summation of the ``$a$ steps" by $\alpha$, i.e. $\alpha=\sum_i a_i$, and thus we characterize the $\gamma$ parameter of an infected node dependent on the specific $\alpha$ value of the node, as:
\begin{equation}
    \label{eq:new_indect}
        \gamma(\alpha) = \gamma_0\xi^\alpha\;.
\end{equation}

\section{\label{sec:tree}Tree graphs}
We start our analytical and numerical work with the case of mutated pathogens spreading on a simple binary rooted tree graphs or a 3-regular Bethe lattice. Binary trees are trees in which each node has exactly two offspring. Binary trees and trees in general have a convenient structure for studying epidemic spreading with multiple pathogens, since each node can be infected by only one other node, its parent, pathogens have only one infection path to each node. This structure excludes the possibility for one pathogen to bypass other pathogens and infect new nodes via a different path. As a result, a new infected node always has the same number of susceptible neighbors (its offspring), unlike other networks which will be discussed below.

In order to solve the epidemic growth of the tree, we start by computing the number of infected nodes in each layer of the tree (where a layer can be described by the distance from the tree's root). Since after few infections along an infection path the initial pathogen is likely to produce multiple different pathogens, we compute the number of infected nodes for each pathogen and consider their sum to get the total infected nodes. Since all the nodes have a constant number of two offspring (or two susceptible neighbors) and the probability for a new infected node of having a higher $\gamma$ or having a lower $\gamma$ is $1/2$, 
we can write a recursive equation for the mean number of nodes in a specific layer $d$ that are infected by a specific pathogen $\alpha$ as:
\begin{equation}
    \label{eq:I}
        I_{\alpha}^d = \frac{1}{2}\cdot2P_{\alpha-1}I_{\alpha-1}^{d-1} + \frac{1}{2}\cdot2P_{\alpha+1}I_{\alpha+1}^{d-1}\;,
\end{equation}
where $P_\alpha$ is the probability for an infected node with pathogen $\alpha$ to infect one of its offspring and $d$ is the distance of the specific layer from the tree's root. Since the $\alpha$ possible values are: $-(d-1)<\alpha<(d-1)$, the mean total number of infected nodes with distance $d$ from the root is:
\begin{equation}
    \label{eq:I_perD}
        I^d = \sum_{\alpha=-(d-1)}^{d-1}\left({P_{\alpha-1}I_{\alpha-1}^{d-1} + P_{\alpha+1}I_{\alpha+1}^{d-1}}\right)\;.
\end{equation}

The probability of a specific node with a pathogen $\alpha$ to infect its neighbors, $P_\alpha$, is also a function of the other parameters of the epidemics -- $\lambda$ and $r$. Considering that an infection occurs when the Poisson process of infection (with mean time $\lambda$) is activated before the first of the recovery Poisson process (with mean time $r$) and the death Poisson process (with mean time $\gamma(\alpha)$) is activated, $P_\alpha$ is calculated as follows,
\begin{eqnarray}
    \label{eq:probability}
        P_{\alpha} &=& \int_{0}^{\infty}\left(\frac{1}{\gamma(\alpha)}+\frac{1}{r}\right) \exp \left(-\left(\frac{1}{\gamma(\alpha)}+\frac{1}{r}\right)t\right) \,dt\ \times \nonumber\\ &&\int_{0}^{t}\frac{1}{\lambda}\exp\left(-\frac{1}{\lambda}\tau\right)\,d\tau\;,
\end{eqnarray}
The solution of the integral is:
\begin{equation}
    \label{eq:solveint}
        P_{\alpha} =  \frac{r\gamma(\alpha)}{\gamma(\alpha)(r+\lambda)+r\lambda}\;.
\end{equation}
The solution of the integral yields the probability of infection for each pathogen. Eq. (\ref{eq:solveint}) agrees with the intuition, in that pathogens with a longer mean lifetime are more likely to infect their susceptible neighbors. Respectively, as time goes on we expect both the number of mutated pathogens, and the values of $\gamma(\alpha)$ of each pathogen to increase.

We substitute Eq. (\ref{eq:solveint}) in Eq. (\ref{eq:I}) and set the pathogen population in each layer to be
\begin{equation}
    \label{eq:pip_I}
        I_{\alpha}^d = r\gamma_0\xi^\alpha
        \left(\frac{I_{\alpha-1}^{d-1}}{r\gamma_0\xi^\alpha(r+\lambda)+\xi r\lambda} +\frac{I_{\alpha+1}^{d-1}}{r\gamma_0\xi^\alpha(r+\lambda)+\frac{1}{\xi}r \lambda}\right)\;.
\end{equation}
A substitution of Eq. (\ref{eq:pip_I}) in Eq. (\ref{eq:I_perD}) gives the solution for the total number of infected nodes in each layer of the tree.

In addition, the infection probability $P_\alpha$ (Eq. (\ref{eq:solveint})) is employed to calculate the basic reproduction number, also known as $R_0$, for each pathogen. Since binary trees have a constant number of two susceptible neighbors for each node, the basic reproduction number is independent of the location of the node in the tree and depends only on the pathogen, as follows
\begin{equation}
    \label{eq:r0solve}
        R_{0\alpha} = 2P_\alpha = 2\frac{r\gamma_0\xi^\alpha}{\gamma_0\xi^\alpha(r+\lambda)+r\lambda}\;,
\end{equation}
resulting in the epidemic $\gamma$--threshold for mutated pathogen being $\gamma_{TH}=\frac{r\lambda}{r-\lambda}$. 
We can also use this calculation to generalize the equation to tree structures with different offspring number. The only adjustment required is to set the number of offspring as a factor in Eq. (\ref{eq:r0solve}).

We note here that in this work we model the case where the recovery mean time $r$ is greater than the infection mean time $\lambda$ only. This guarantees the existence of the threshold $\gamma_{TH}$ that is mentioned above, in binary trees, such that pathogens with transmissibility below the critical threshold can infect a finite number of nodes only when spreading in a binary tree \cite{TH_epidemics_net,2pathogensNewman}, while pathogens above this limit infect an infinite number of nodes (in the infinite tree case, or a finite fraction of the nodes in a finite tree case) with positive probability.

In the case of a mutating pathogen, we can show that if there is a threshold for the spreading process, then even with an initial pathogen with parameters far below the threshold, there is still a positive probability for the epidemic to infect an infinite number of individuals. Consider a tree with an epidemic threshold $\gamma_\mathrm{TH} = X$. The initial pathogen is described with $\gamma_{0} = X\xi^{-m}$. Now, we would like to show that with positive probability the initial pathogen can create a fitter mutation that will surpass the epidemic threshold. Since every new infected node has a probability of $\frac{1}{2}$ of gaining a longer mortality mean-time parameter than the infecting node, and the probability of infection is known for each pathogen (Eq. (\ref{eq:solveint})), we can compute the probability of surpassing the epidemic threshold in exactly $m$ steps as follows:
\begin{equation}
    \label{eq:m_Steps}
       \frac{1}{2^m}\prod_{s=1}^{m} P_{-m+s}\;,
\end{equation}
where $P_{-m+s}$ is calculated by Eq. (\ref{eq:solveint}).

Therefore, since $P_{-m+s} \geq 0$ for any $m$ and $s$, it follows that the probability of the epidemic to reach an endemic state is always positive, regardless of the initial mortality rate.

In fact, the total probability to get the threshold point is considerably greater than the bound presented above, since the pathogen can evolve in more than $m$ steps and with more then one optional path. Therefore, we can say that the total probability to have a new mutated pathogen above the epidemic threshold during the spreading process is greater than zero for any initial parameter value. As a result, we can say with certainty that for any initial mutated pathogen, there are outcomes where the epidemic reaches an endemic state, regardless of the initial parameters' value.

In order to confirm our calculations we performed numerical simulations of mutating pathogens in trees. The simulations start with an initial infected node (also described as the tree's root), with the initial pathogen $\alpha = 0$, or $\gamma = \gamma_0$. The node is also set with the constant parameters $r=r_0$, $\lambda = \lambda_0$. The pathogen parameters are used to simulate Poisson processes for the pathogen's life time and for each infection time to each of its offspring. If an offspring is infected, we create it as an infected node with $\gamma_{new} = \gamma_0\xi^{\pm1}$. In addition, we set to it the constant parameters -- $r_0$ and $\lambda_0$ -- for the spread process. This is repeated iteratively, step by step, until the end of the epidemic. We define a stopping condition when there are no ongoing infection processes.

In order to obtain more interesting results, the simulation should begin when the initial pathogen's parameters are around the critical value of the reproduction number $R_0=1$. Accordingly, based on Eq. (\ref{eq:r0solve}), we set the initial parameters to be around the conditions below:
\begin{equation}
    \label{eq:THcondition}
        \gamma_0(r+\lambda)=r\lambda\;,
\end{equation}
where $\alpha = 0$.

\begin{figure}
\includegraphics[scale=0.65]{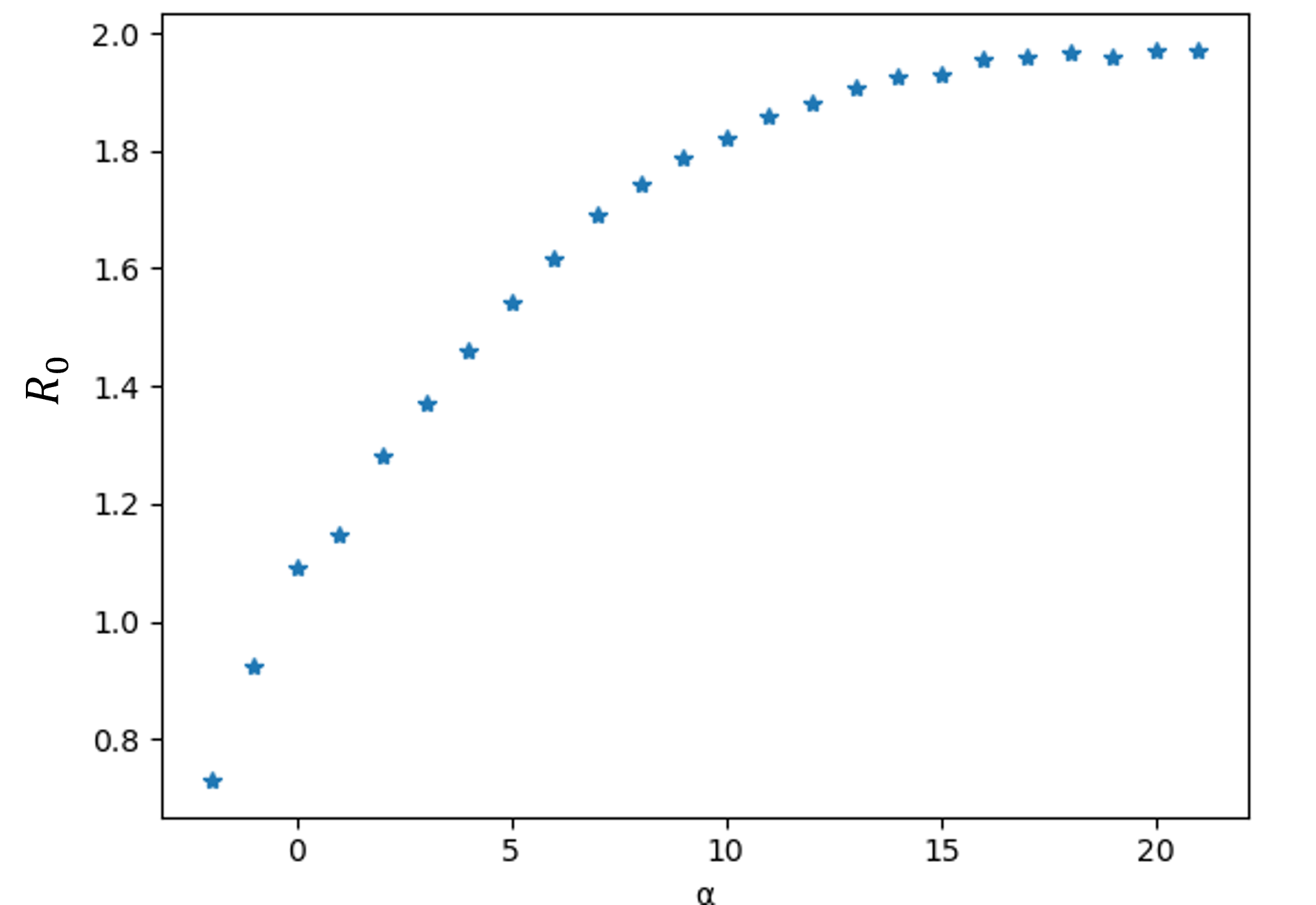}
\caption{\label{R by alpha eps 1.25} The basic reproduction Number ($R_0$) as a function of $\alpha$, using the following parameter values: $\gamma_0=2.25$, $\lambda = 2$, $r = 20 $ and $\xi=1.25$, based on simulations results.}
\end{figure}

\begin{figure}
\includegraphics[scale=0.44]{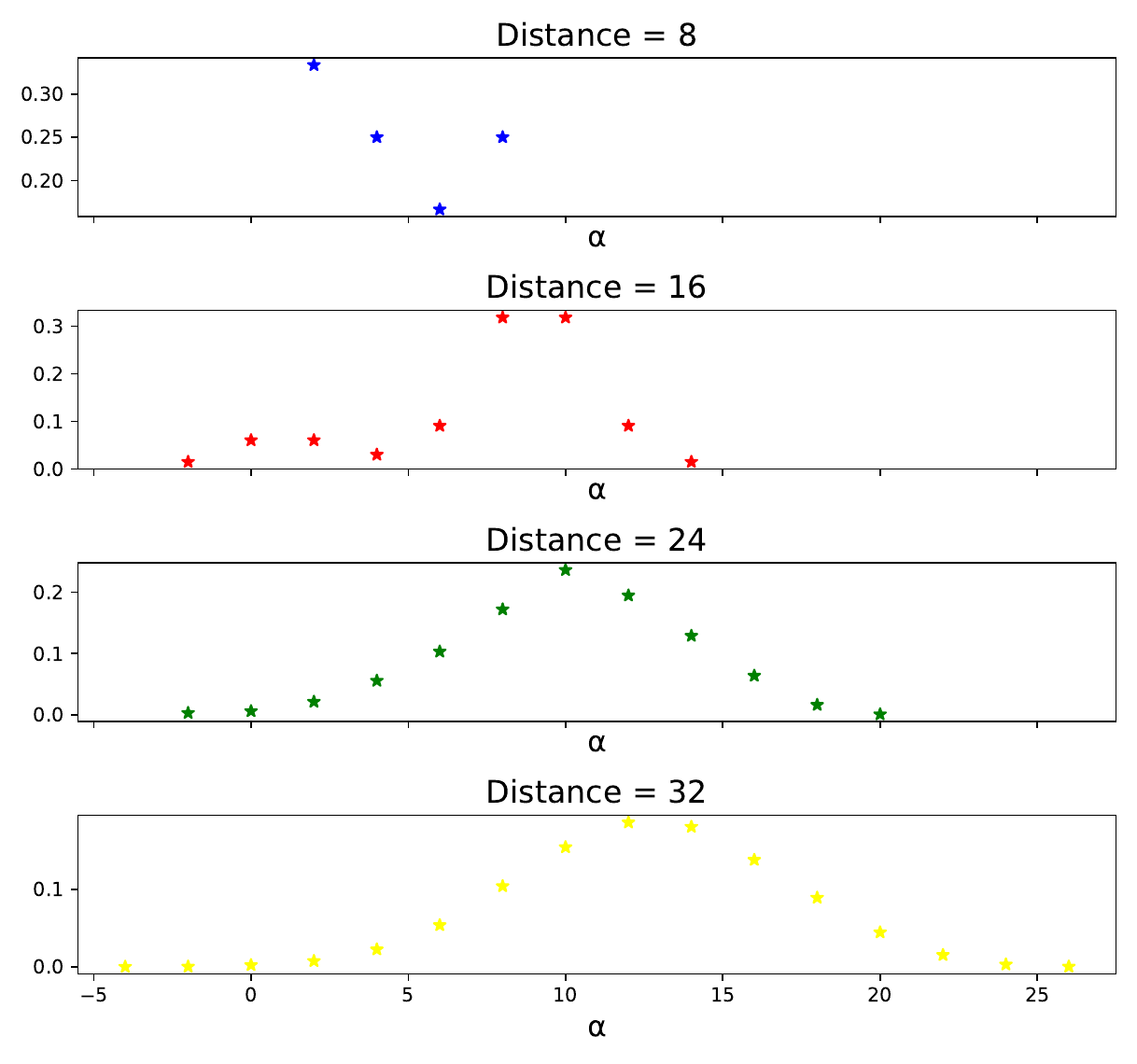}
\caption{\label{alpha distribution} The pathogens distribution on nodes with different distances from the root node (values on titles). Initial parameters are as in Figure \ref{R by alpha eps 1.25}. The $x$-axis is $\alpha$ and the $y$-axis is the ratio of the number of nodes infected with each pathogen to the total number of infected nodes in this distance.}
\end{figure}

We present our simulations results in tree graphs in Figs. \ref{R by alpha eps 1.25} and \ref{alpha distribution}. Figure \ref{R by alpha eps 1.25} shows an excellent fit with the solution of Eq. (\ref{eq:r0solve}), where $R_0$ grows as a function of $\alpha$, with upper bound in $R_0=2$ due the graph's structure (since every node has two offspring). This agrees with our intuition and the epidemiological logic, in that a longer mean mortality period increases the probability to infect nearby susceptible nodes, resulting in an increase of $R_0$.

Figure \ref{alpha distribution} demonstrates how higher values of $\alpha$ (and $\gamma$ as a result, since $\gamma$ is dependent on $\alpha$, see equation \ref{eq:new_indect}) impact the spread of pathogens in the network, where pathogens with higher values of $\alpha$ (or $\gamma$) gain more and more dominance over the network, when they infect a greater fraction of the new infected offspring. These results are in good agreement with Eq. (\ref{eq:pip_I}), where pathogens with a greater $\gamma$ (and a greater $\alpha$ respectively) are more likely to infect new nodes. It should be noted that the right tail in each scatter diagram in which the proportion of pathogens with large $\alpha$ seemingly decreases, is due to the finite time of the simulations which does not enable these pathogens, appearing in later stages of the simulations, to infect a significant portion of susceptible nodes. In any case, the principle of proportional growth with increasing $\alpha$ is conserved, and as time goes on we are more likely to see pathogens with higher values of $\gamma(\alpha)$.
\section{\label{sec:er}Random networks}
Random networks are networks with randomly generated edges. Random networks have been thoroughly studied over several decades and have been used as a basis for many studies on the field of complex systems in general and complex networks in particular \cite{PhysRevE_Random_newman}. The most familiar random network model is the Erdős–Rényi model ($G(n,p)$) \cite{erdos_re}.

It has been shown that even in large random networks, properties of small world networks remain \cite{random_dis}, where the average distance between nodes in the network is $d\sim ln(N)$. These structural properties enable epidemics to spread in the network with only a small number of steps from any initial infected node to any other node in the network. As a result, mutating pathogens generally create only a limited number of mutations. 

In order to model the epidemic spreading on random graphs, we divide the spreading process into two parts -- short and long range spreading. At the beginning of the epidemic spread, when most of the nodes in the graph are not yet infected, the random graph can be assumed to behave as a tree graph, which means that each new infected node is free to infect all its neighbors except its source of infection, and ``bypasses'' do not yet exist. Therefore, the equation for the basic reproduction number on the short time spreading can be written as in the tree section, with the necessary changes for random graph structure:
\begin{equation}
    \label{eq:R_Random}
        R_{0\alpha} = \frac{r\gamma(\alpha)}{r\gamma(\alpha)+\lambda(r+\lambda)} <k>\;,
\end{equation}
where $<k>$ is the expectation of the number of links of each node in the network. When a significant part of the nodes in the network have already been infected (long range spreading part), some of the neighbors of an infected node are already infected, and the number of susceptible neighbors no longer behaves exactly like the graph degree distribution. 

For the numerical part of this section, we performed numerical simulations similar to those in the previous section. However, in order to simulate the process for random networks, we first generated random networks based on Erdős–Rényi model and simulated the spreading process on them. The simulation starts with a random initial infected node in the network. The initial node receives an initial pathogen with properties around the critical epidemic threshold value based on the graph structure. The simulation initiates Poisson processes of the infection process for its neighbors as well as the recovery/death of the infected node itself. The Poisson processes are simulated based on the pathogen's parameters as their average times characteristics. When another node becomes infected, it receives a new pathogen with the same constant parameters ($r$, $\lambda$) as its source node, and with a new mutating parameter ($\gamma$) according to Eq. (\ref{eq:new_indect}). Afterwards, similar Poisson processes are initiated for the new infected node, but now with the new pathogen's mortality parameter. The simulation also considers the status of the node's neighbors, where possibly only some of them are susceptible and could be infected based on the Poisson process. The simulation continues until it stops when most of the network is no longer susceptible, or until we count several simulated days with no new infections on the network. 

\begin{figure}
\includegraphics[scale=0.43]{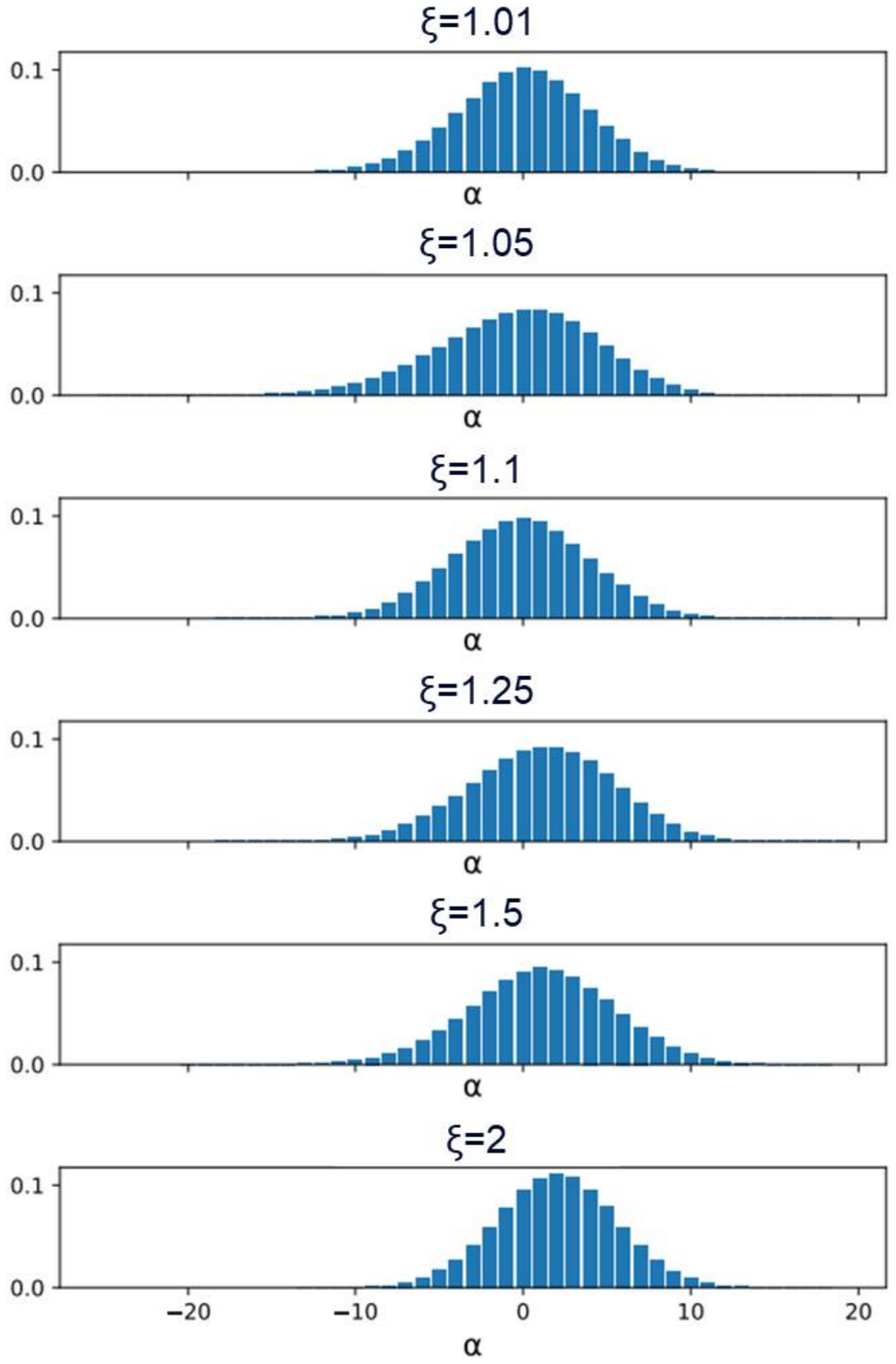}
\caption{\label{Random- alpha distrub} The pathogen distribution when most of the network's nodes have already been infected. Each sub graph presents a simulation with a different $\xi$ value (see figures titles). the $x$-axis is the $\alpha$ value, and the $y$-axis is the proportion of nodes with that pathogen.}
\end{figure}

In Fig. \ref{Random- alpha distrub} we present the distribution of the pathogens in the network after a long spreading time, when most of the network nodes are already not susceptible. We expect the resulting distribution to be similar to the normal distribution (since we can partly describe the mutation process as a random walk), with a slight tendency for the positive side inasmuch as pathogens with positive $\alpha$ (or larger $\gamma$) are likely to have a greater portion of the network nodes (see Eq. (\ref{eq:solveint})). This tendency is affected by the $\xi$ value, where greater values yield greater mutations and advantages to the mutated pathogens with greater $\alpha$ as we can see also in the figure.

\begin{figure}
\includegraphics[scale=0.5]{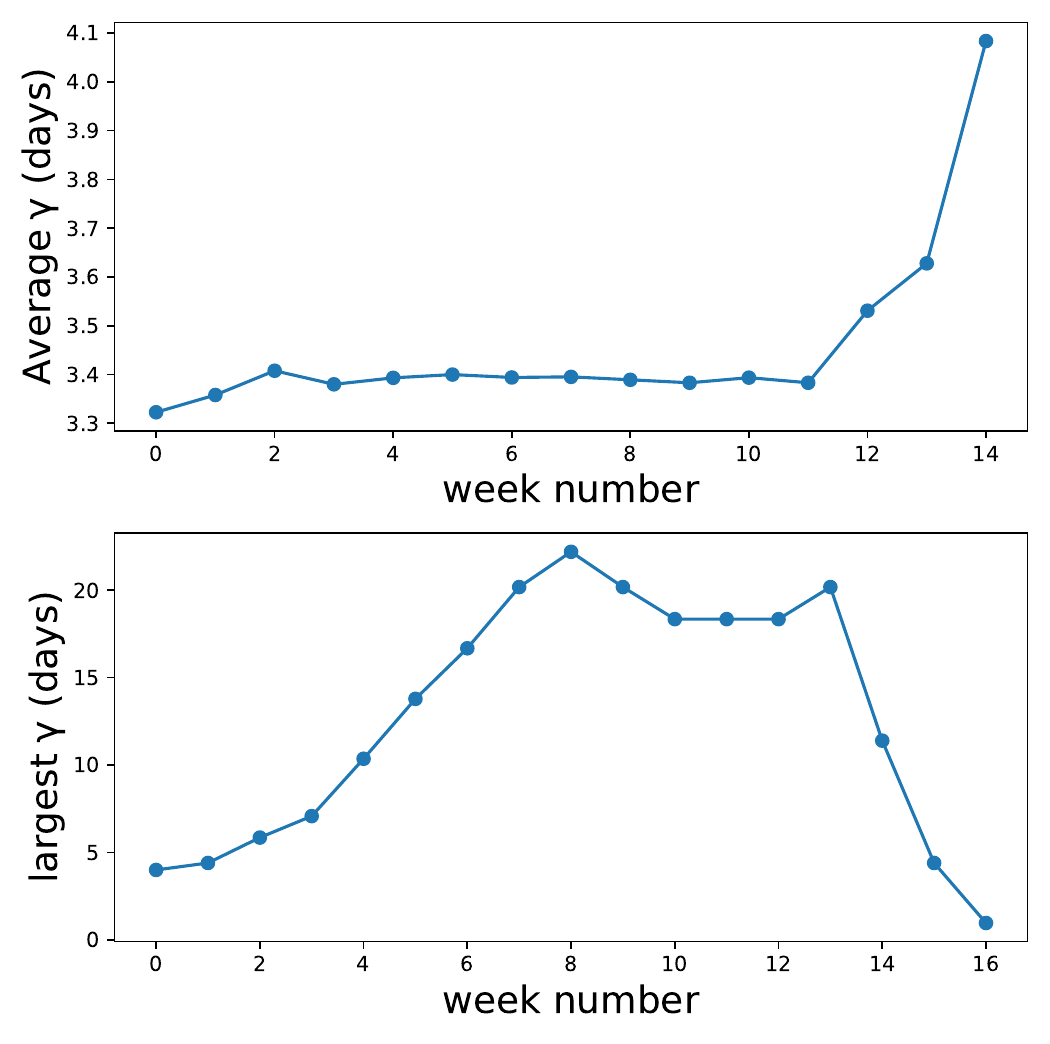}
\caption{\label{Random- larger and average} The evolution of a mutating pathogen during the epidemic spreading, on a random network with 2M nodes. the initial parameter is set around the epidemic threshold with $\xi=1.1$. The upper sub figure shows the average value of the mutating pathogens' $\gamma$ in nodes which were infected in each week, while the lower sub figure shows the largest value of $\gamma$ in each week. The $x$-axis represents the week number on the simulation, and the $y$-axis is the $\gamma$ value (in days).}
\end{figure}

We also studied the evolution of the mutated pathogen during the time it spreads, as presented in Fig. \ref{Random- larger and average}. In this figure, it is shown how pathogens with greater values of $\gamma$ appear during the epidemic spread. The figure presents the average and the greatest value of $\gamma$ of the infections in a specific simulated week. The pathogens evolution can be seen in the bottom figure, where for each week during the simulations the greatest $\gamma$ pathogen that appeared during that week is presented. Then, after some time, the pathogens with the greater values infect more new nodes, and the average $\gamma$ value increases respectively (top figure). At the same time, the great $\gamma$ pathogens (as well as the other pathogens) stop infecting nodes, as a result of the network structure and its finite size, limiting the number of infection steps, which also limits the pathogen's mutation.

\section{\label{sec:sf}Scale free and real world networks}
Scale free networks are networks with a power-law degree distribution. In particular, it means that in this graph most of the nodes have low degrees, while a small fraction of nodes get very high degrees. These nodes are sometimes denoted as ``hubs'' due to the role they play in the network.

Over the last decades, the scale free network model was found to fit well many real world networks, such as the WWW, the Internet, air travel networks and social networks \cite{Redner-eurp-phs-j-1998,albert-nature-1999,Barabasi-science-1999,Newman-pnas-2001}.

Scale free networks have a strong ``small world effect'', since the average distance in these networks is very short and scales as $d \sim ln ln N $ \cite{ultrasmallworld}. This makes the average ``infection path'' (the number of infections needed to infect a far node in the network) extremely short, so the pathogen does not undergo enough infection steps in order to create large mutations.

In this section we focus on numerical simulations. To generate the simulated networks, we used the Barabási–Albert model \cite{ba_model,BARABASI2002590}. We also worked with real-world data, from open sources.

\subsection{\label{sec:sim_sf}Simulated scale free networks}

\begin{figure}
\includegraphics[scale=0.5]{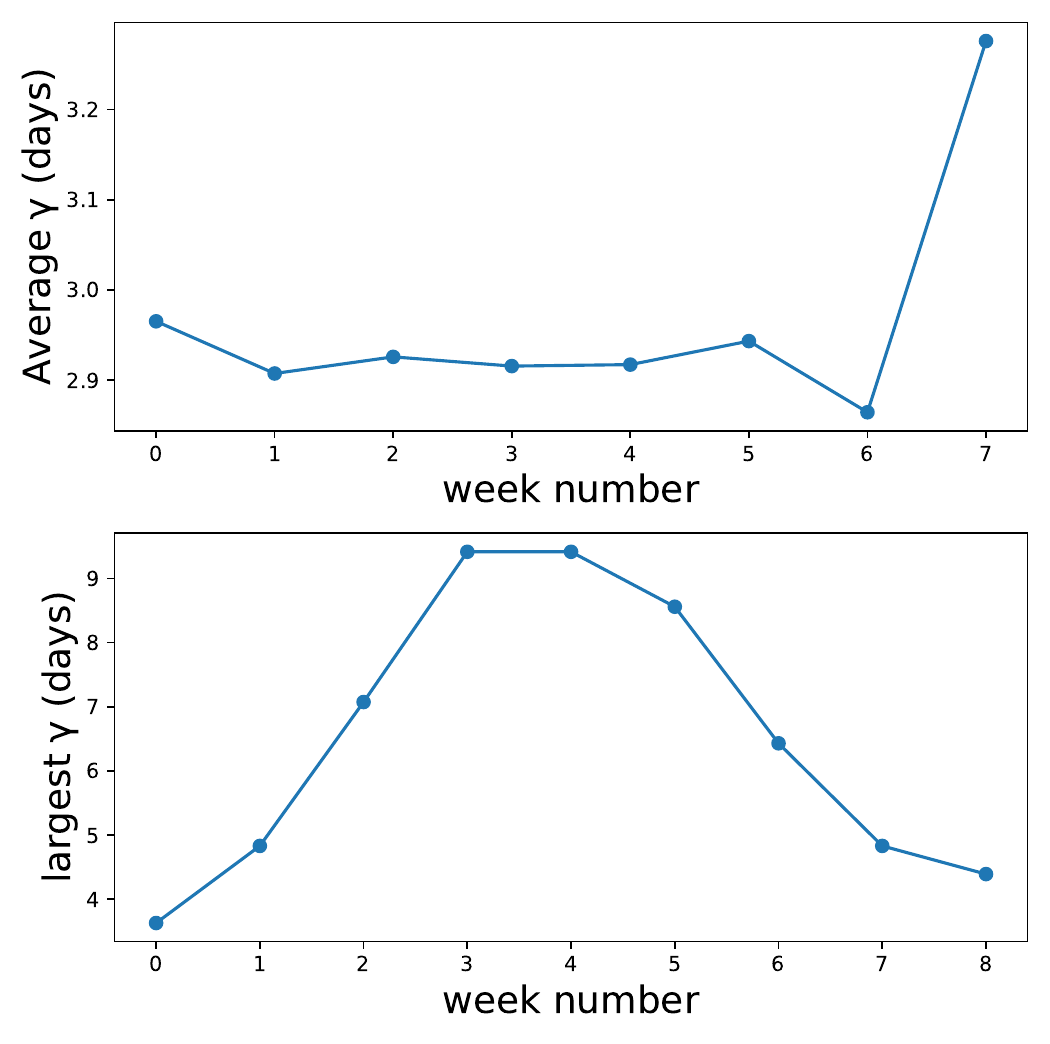}
\caption{\label{SF- gamma max and mean} The evolution of the mutated pathogen (similar to Fig. \ref{Random- larger and average}) during its spreading in a scale free network, with $\xi=1.1$. The initial pathogen's parameters are set around the epidemic threshold (which is determined by the network's finite size).}
\end{figure}

As expected from the network structure, Fig. \ref{SF- gamma max and mean}  shows how scale free networks are infected faster (if we compare to random networks with the same initial conditions as those presented in Fig. \ref{Random- larger and average}) and are less likely to induce large mutating. Since paths between nodes in scale free networks are likely to be much shorter, there are not enough infection events along a path to create a pathogens with a mutated parameter which is considerably larger than the initial pathogen. Indeed, as we can see in the top figure of Fig. \ref{SF- gamma max and mean}, the average pathogen's mortality parameter is almost constant (the growth at the end of the spreading process is only about 12\%, similar to only one mutating step, since $\epsilon$ here is 1.1).

\subsection{\label{sec:realworld}Social network}
In order to study the  mutated pathogen model under more realistic conditions, we used the network of Deezer Europe social network \cite{feather}. In this network the nodes are Deezer users from European countries and edges are mutual follower relationships between them.

\begin{figure}
\includegraphics[scale=0.54]{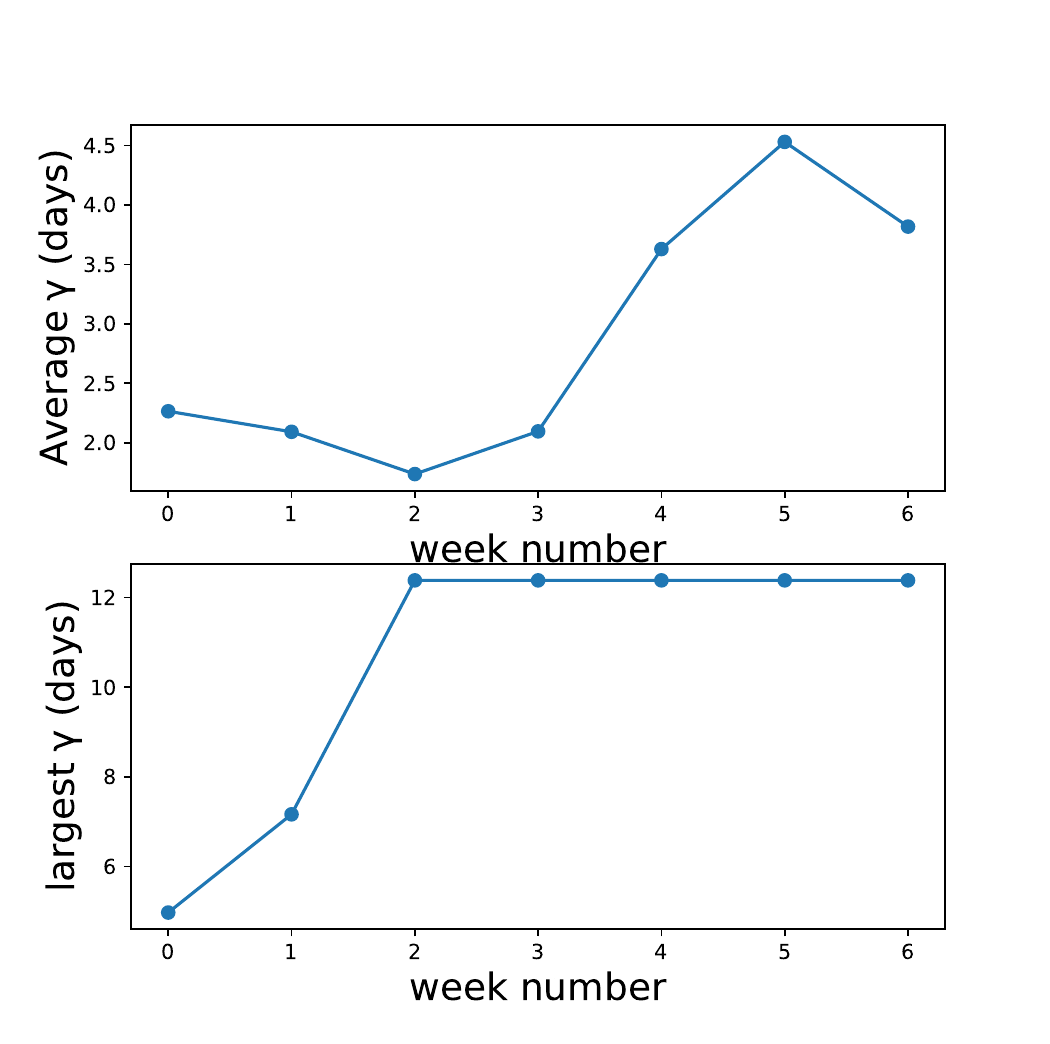}
\caption{\label{deezer pathogen evolution} The evolution of a mutated pathogen (similar to Figs. \ref{Random- larger and average},\ref{SF- gamma max and mean}) during its spreading in the Deezer European social network, with $\xi=1.2$. The initial pathogen's parameters are set around the epidemic threshold, based on the network structure.}
\end{figure}

As we expected, the results of the simulation using the Deezer Europe social network are similar to results for the generated scale free networks, as seen in Figure \ref{deezer pathogen evolution}. The pathogen begins its evolution when a few nodes are infected with pathogens having higher value of $\gamma$, that after some time get dominance over the network, and the average pathogen's value of $\gamma$ is growing. Since scale free networks have a very small diameter, which is almost independent of the size, the average distance in the Deezer netwrok is very close to the average on our generated scale free network even though it is much smaller. Therefore, the total infection time is pretty similar between them, while random networks of these sizes will present a larger difference in the spreading time.

\section{\label{sec:grid}square grid graph}
Square grid graph is a type of a lattice, or a grid graph, which is a cyclical graph (considering an infinite graph or a graph with periodic boundary conditions) in which each node has four neighbors that are identical in structure (i.e., it is vertex transitive).

Our primary motivation for investigating this kind of graphs is the large average distance they possess. In addition, the square grid graph, as small world networks and unlike tree graphs, can have more than one path between two nodes. Additionally, unlike random graphs, grids contain a large number of small cycles. This yields the effect of competing paths, by which competing pathogens can ``get around'' other pathogens and become dominant in the network. These structural properties allow the mutating effect to be more strongly pronounced, and make this structure interesting for our study.

\begin{figure}
\includegraphics[scale=0.55]{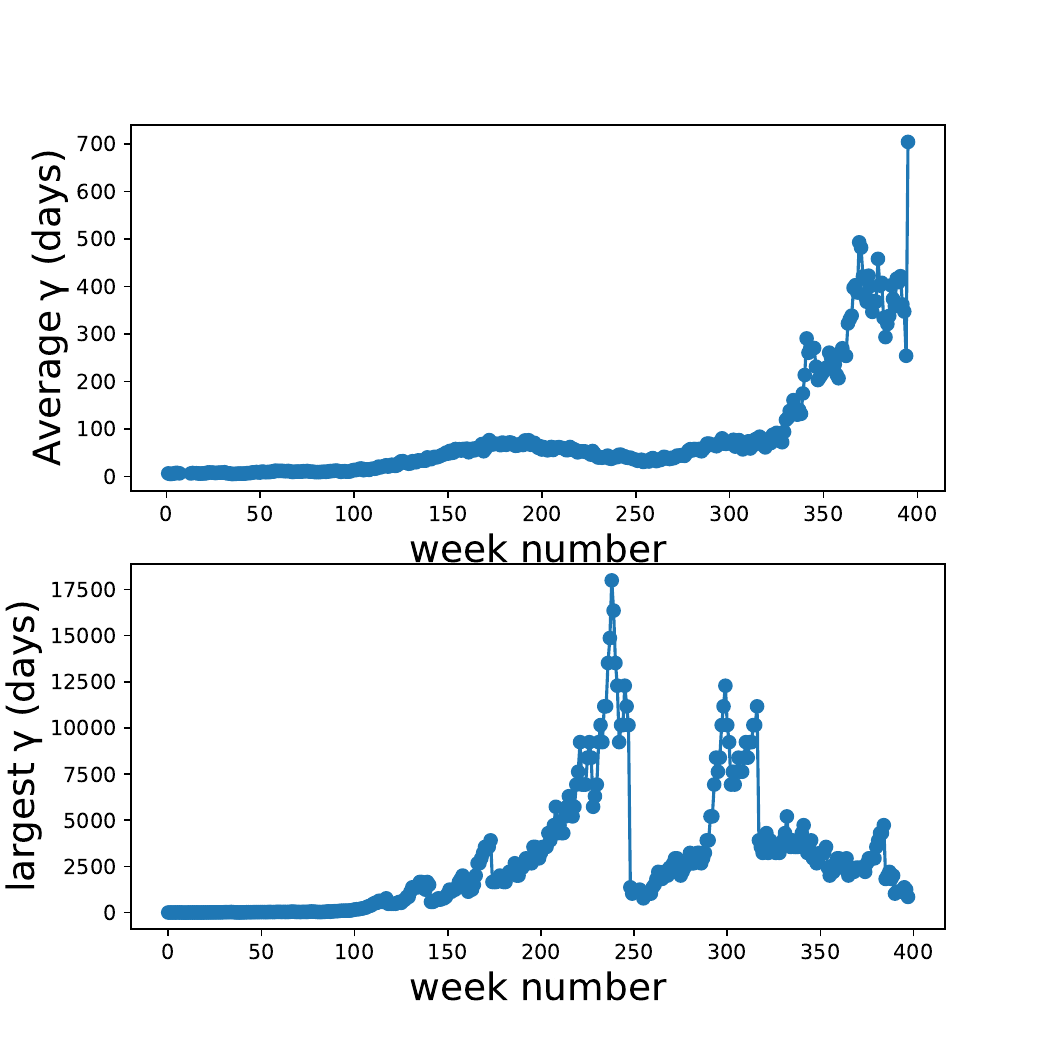}
\caption{\label{grid gamma} The evolution of the mutated pathogen in a square grid graph with $500\times 500$ nodes. The initial parameter is set around the epidemic threshold, and $\xi = 1.1$. }
\end{figure}

Figure \ref{grid gamma} shows how the mutating parameter gets greater over time in a square grid graph, where only the fittest pathogens survive after a long path. It seems that at some point most of the low pathogens disappear, which makes the average parameter grow much faster.

\begin{figure}[th]
\includegraphics[scale=0.63]{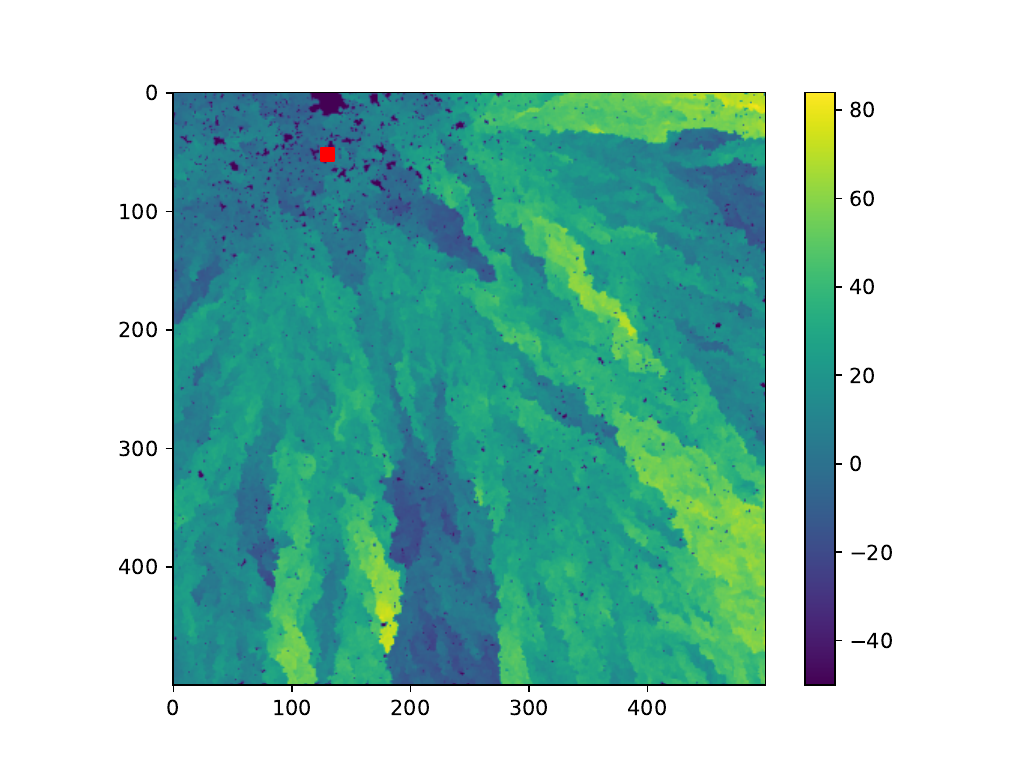}
\caption{\label{alpha heatmap} The heat map for $\alpha$ values in  a square grid graph with $500\times 500$ nodes. In case of a susceptible node, we set a very low $\alpha$ value (dark point). The initial infected nodes are colored in red.}
\end{figure}

Another advantage of the square grid graph is the ability to easily visualize the graph's structure and better understand the flow of the epidemic in it.  Figure \ref{alpha heatmap} displays the mutation of the pathogen during the epidemic spreading in the network. The $\alpha$ value of each node is presented as a color in the figure. Dark pixels represent never infected susceptible nodes. The initial infected node is colored in red. It can be seen that close to the initial node where pathogens are likely to have smaller value of $\gamma$, more nodes remain susceptible, i.e. the infection probability is lower. As the distance from the initial node increases, these holes become less and less common. That is, after the pathogen mutates the infection probability becomes larger (due to the decreased mortality, leading to higher transmissibility values) and thus far nodes have a much lower probability to stay healthy without getting infected. At the same time, farther  from the initial node we can see much higher variance in the parameter value. This can be explained by what biologists define as the ``founder affect'', when in each area the infection can be envisioned as a tree with its founder as a root. The parameter value for this ``founder'' node is likely to have a large effect on the properties of pathogens existing in the area around it. Thus, one can see that in the figure there are patches of brighter and darker colors, which are usually brighter than the vicinity of original infection, but have a large variation that is due to the exact parameter value for the pathogens that arrived at this direction.  
\section{\label{sec:tree}Test case -- The Covid-19 pandemic}
We tested our theory and predictions on the real well-known `Corona' pandemic. This pandemic originated in Wuhan, China in December 2019. The virus responsible for the global break is named SARS-CoV-2 (Severe Acute Respiratory Syndrome Coronavirus 2). Infection with this virus caused the disease known as COVID--19 (Corona Virus Disease 2019). The World Health Organization (WHO) declared The `corona' a global pandemic in March 2020 \cite{WHO-website1}. From December 2019, as mentioned, the pandemic continued until May 2023 when the WHO ended the Covid--19 global health emergency \cite{WHO-website2}. Throughout the pandemic, five variants of concern (VOC) of Covid-19 emerged, each of them with its own special characteristics -- Alpha, Beta, Gamma, Delta and Omicron. Of the five VOCs, three of them -- Alpha, Delta and Omicron -- were very infective and transmissible, and shortly after they emerged they reached dominance over other variants that coexisted with them and caused a global pandemic wave. The Alpha variant first emerged in September 2020. It was the first VOC, and appeared after the first strain of Covid-19 variants that originated in Wuhan and referred as to ``wild-type'' strain. This variant became dominant over the ``wild-type'' strain, and caused the first world pandemic wave that occurred from December 2020 until early 2021. The Delta variant first emerged in December 2020. This variant reached dominance over the alpha variant and other variants coexisted with it, and caused the second pandemic wave over most of 2021 \cite{WHO-website3,ECDC-website}. The Omicron variant appeared in November 2021. This variant has a very high transmissibility, and shortly after it emerged it took the dominance over the Delta and other variants and caused the third pandemic wave that took place in 2022-2023. The other two VOCs -- Beta and Gamma -- emerged first in May 2020 and November 2020, respectively. Each of them showed specific properties that made them riskier than variants strains (such as high transmissibility, severe disease, etc.) that existed before they appeared, and for this reason they were designated as VOCs \cite{Raddad-cln-inf-dis-2022,harrigan-JMIR-2024}. 

We applied our theory using real epidemiological measurements of the infection, mortality and recovery mean time of each of the Covid-19 VOCs Alpha, Beta, Delta and Omicron. 

The following is a detailed explanation of how we measured the three mean times. In general, to measure these mean times we used some other epidemiological auxiliary measures related to the Covid-19 VOCs, as will be detailed below. These auxiliary measures have been evaluated in many studies and, naturally, their values vary in different studies. Therefore, whenever the evaluation of a measure was different between studies, we consider its value as the mean of the various values that were presented. Here is the way we measured each of three mean times -- infection, mortality and recovery:

(i) Measurement of infection mean time: In epidemiological terms, infection mean time is referred to as the measure named \textit{generation time} -- the interval between when an individual becomes infected (primary infection) and when that individual transmits the infection to another (secondary infection). Since there are difficulties in measuring the generation time, it is replaced by the \textit{serial interval} measure, that is the interval between the appearance of symptoms of the primary infection to the symptoms appearance of the secondary infection \cite{madewe-BMC-2023,chen-naturecom-2022}. For the Covid-19 VOCs, the serial interval was measured to be as follows: Alpha -- 5.2 days \cite{madewe-BMC-2023,jusot-4open-2022}, Beta -- 4.5 days \cite{xu-BMCmed-2023}, Delta -- 3.7 days \cite{madewe-BMC-2023,xu-BMCmed-2023,du-jrnl-trvl-mdcn-2022}, and Omicron -- 3 days \cite{du-jrnl-trvl-mdcn-2022,madewe-BMC-2023,niu-front-pub-hlth-2022}.

(ii) Measurement of mortality mean time: The mortality mean time which is the mean interval between infection and death, has not been directly studied and evaluated for the Covid-19 VOCs. Therefore, for measuring it we split this interval into two parts that have been studied and measured -- the interval between infection and the onset of symptoms known as the \textit{incubation period}, and the interval between hospitalization due to symptoms and death for individuals who died from the disease. We also considered 1-2 days between the onset of symptoms and hospitalization. For each of the Covid-19 VOCs, the combination of the incubation period and the hospitalization-to-death interval plus 1-2 days between them gives a measure of its mortality mean time. For the Covid-19 VOCs, the incubation period was measured to be as follows: Alpah -- 4.5 days \cite{xu-BMCmed-2023,ward-epdm-and-inf-2023,manica-epdm-and-inf-2023}, Beta -- 4.5 days \cite{wu-JAMA-2022,roe-heliyon-2022,desai-pediat-rsrch}, Delta -- 4.35 days \cite{manica-epdm-and-inf-2023,du-jrnl-trvl-mdcn-2022,ward-epdm-and-inf-2023} and Omicron -- 3.35 days \cite{ao-MedComn-2022,jansen-MMWR-2021,du-jrnl-trvl-mdcn-2022}. The hospitalization-to-death time is: Alpha -- 14.3 days, Beta -- 4 days, Delta -- 12.8 days and Omicron - 16 days \cite{ward-epdm-and-inf-2023,hirachund-safrica-fmly-2023}. Combining the previous data gives the following measurements for the VOCs mortality mean time: alpha -- 20.3 days, Beta -- 10 days, Delta -- 18.7 days and Omicron -- 20.9 days.

(iii) Measurement of recovery mean time: As in the mortality mean time, the recovery mean time, which is the mean interval between infection and recovery, has not been widely studied and directly evaluated. For measuring the recovery mean time, we split this interval, as before, into the incubation period and the time interval from hospitalization due to symptoms to recovery for individuals who recovered from the disease (plus 1-2 days between symptoms onset and hospitalization). The combination of the incubation period and the hospitalization-to-recovery interval plus 1-2 days between them gives a measure of the recovery mean time. For Covid-19 VOCs, the hospitalization-to-recovery time was measured as follows: Alpha -- 11 days, Beta -- 6 days, Delta - 6 days, Omicron -- 3 days \cite{goga-MedRxiv-2021,gareeva-tera-arxv-2024}. Combining these data with the incubation period detailed above yields measurements of the recovery mean time of Covid-19 VOCs as follows: Alpha -- 17 days, Beta -- 12 days, Delta -- 11.9 days and Omicron -- 7.9 days.

In Table \ref{tab:table1} we summarize the various measurements for the VOCs as calculated before. Each of the Alpha, Beta, Delta and Omicron variants is represented in a specific raw in the table. The first three columns list the infection mean time ($\lambda$), the mortality mean time ($\gamma$) and the recovery mean time ($r$), respectively. In the fourth column, the infection probability $P_{inf}$ is calculated for each of the variants according to the formula we developed in Eq. (\ref{eq:solveint}), dependent on the three mean times $\lambda$, $\gamma$ and $r$. The fifth column is devoted to another important parameter that is the mortality probability $P_{mort}$. This probability refers to an event in which an infected individual dies from the disease. In our model, it occurs when the inter arrival time of the mortality Poisson process occurs before the inter arrival time of the recovery Poisson process. Considering that the inter arrival time of a Poisson process is exponentially distributed and the immemorial of the exponential distribution, we get for the mortality probability as follows: $P_{mort}=\frac{1/\gamma}{1/\gamma+1/r}$. In the fifth column of the table we calculated $P_{mort}$ for each of the VOCs applying this formula.

\setlength{\tabcolsep}{12pt} 
\begin{table*}[ht]
  \caption{Summary of mean time intervals in the course of infection by variant.}
  \label{tab:table1}
  \centering
  \begin{tabular}{l c c c c c}
    \toprule
    Variant & Infection mean time ($\lambda$) & Mortality mean time ($\gamma$) & Recovery mean time ($r$)  & $P_{inf}$  & $P_{mort}$\\
    \midrule
    Alpha   & 5.2 & 20.3 & 17 &0.64 & 0.46 \\
    Beta   & 4.5  & 10  & 12  & 0.55 & 0.55 \\
    Delta   & 3.7 & 18.7 & 11.9 &0.66 & 0.39 \\
    Omicron & 3 & 20.9 & 7.9 & 0.66 & 0.27 \\
    \bottomrule
  \end{tabular}
\end{table*}

An analysis of Table \ref{tab:table1} leads to several conclusions regarding the spread of the Corona virus in the world population, as follows:

(i) The infection probability $P_{inf}$ of Alpha -- 0.64, Delta -- 0.66 and Omicron -- 0.66, which are the three variants that caused the three waves of the pandemic, is almost equal. Therefore, we deduce that the replacements of the Alpha wave by the Delta wave and of the Delta wave by the Omicron wave did not result from a creation and emergence of a variant with much higher infection probability than the previous one. On the other hand, the explanation for these replacements is found in the infection mean time column. The infection mean time decreased between variants as the pandemic progressed, from 5.2 days for Alpha to 4.5 days for Delta and finally 3 days for Omicron. The infection mean time of a variant indicates how fast the process of infecting individuals with this variant occurs. It is independent of the infection probability. Accordingly, we deduce that the Delta wave replaced the Alpha wave and the Omicron wave replaced the Delta wave, due to the fact that the rate of Delta infections was much higher than the rate of Alpha infections, and the rate of Omicron infections was much higher than that of Delta infections.  

(ii) The mortality probability $P_{mort}$ decreases between the variants Alpha, Delta and Omicron as the pandemic progressed. for Alpha the probability is 0.46, for  Delta it is 0.39 and for Omicron it is 0.27. This fact is an indication to the validity of our main argument that the survivability of a variant increases as the variant virulence decreases. In the Covid-19 case, the dominance reached by Delta over Alpha and by Omicron over Delta, is indeed a result of their increase of the infection rate (as detailed above), but here we identify that factor of low mortality probability has also a significant contribution to the survivability and dominance of a variant.  

(iii) The variant Beta is different from the other VOCs Alpha, Delta and Omicron. Its infection probability 0.55 is much lower than the infection probability of each of the Alpha, Delta and Omicron which is 0.65 approximately. On the other hand, the mortality probability of the Beta variant that is 0.55, is much greater than this probability for each of the Aplha, Delta and Omicron variants. Therefore, in accordance with our theorem, the low infection probability of the Beta variant explains the facts that the Beta did not cause a wave in the epidemic and could not reach dominance over the variants coexisted with it. On the other hand, its infection probability 0.55 is significant and did allowed the Beta to create strains of individuals that were infected with it. Combining this with the Beta's second property of a high mortality probability, gives an explanation why this variant is designated as a VOC. Although its probability of infection was relatively low, it was a very virulence variant such that if an individual was infected with the Beta virus, his chance of dying was very high.

\section{\label{sec:conc}conclusion}
In this paper we studied the phenomenon of continuous mutating pathogens in networked populations. We show that mutating pathogens have a large effect on the way of epidemics spread among populations. We showed that effect over a wide range of networks and specifically we computed the epidemic equations for continuous mutating pathogens on a binary tree graph as well as the basic reproduction number. We also proved that in the case of mutated pathogens that spread in random networks, there is always a positive probability to reach an endemic state, regardless of the  parameters of the initial pathogen. Using numerical simulations, we show how the mutated pathogens spread in many kinds of populated networks and the differences between different structures of networks. The numerical simulations also confirmed the calculations from the analytical parts. The results are consistent with the known effect of the less violent variants that become more dominant over time. In addition, we applied our theory to the case of the Covid-19 pandemic, including analyzing the spreading of the various variants of the primary pathogen over different periods. We provided explanations for the pandemic's development and the variants that emerged and spread in the population.

The behavior of the mutated pathogen in populated networks is yet to be fully understood. Future work can focus on computing the epidemic equations for grid graphs, as grid graphs are more likely to have large mutation effects on the pathogens. One can also consider studying the case of a number of well-connected networks which are interconnected by a limited number of links. This case may be the base to model mutations flows between countries, during the spread of real world epidemics.

\newpage
\section*{\label{sec:ac}Authors contribution}
A.I. contributed to the development of the theory, wrote and implemented the simulations and wrote the draft. R.C. contributed to the theory development and reviewed and edited the draft. A.P. contributed to the development of the theory, wrote the `Test case -- the Covid-19 pandemic' section, reviewed and edited the draft and the presentation and supervised the project of writing the paper.

\nocite{*}

\bibliography{main}

\end{document}